\documentclass{aa}
\usepackage{natbib}
\usepackage{graphicx}
\usepackage{booktabs}
\usepackage{txfonts}

\newcommand{\jmst}{J.~Mol.~Struct.}   

\begin{document}

\title{Magnesium radicals MgC$_5$N and MgC$_6$H in IRC\,+10216
\thanks{Based on observations carried out with the Yebes 40m
radiotelescope at Yebes Observatory, operated by the Spanish
Geographic Institute (IGN, Ministerio de Transportes,
Movilidad y Agenda Urbana).}}


\author{J.~R.~Pardo\inst{1}, C.~Cabezas \inst{1}, J.~P.~Fonfr\'ia\inst{1}, M.~Ag\'undez\inst{1},
  B.~Tercero\inst{2}, P.~de~Vicente\inst{2}, M.~Gu\'elin\inst{3},
  and J.~Cernicharo\inst{1}}
\institute{Instituto de F\'{\i}sica Fundamental, Consejo Superior de Investigaciones
  Cient\'ificas, C/ Serrano 123, 28006 Madrid, Spain \\
  \email{jr.pardo@csic.es,jose.cernicharo@csic.es}
  \and
  Institut de Radioastronomie Millim\'etrique, 300 rue de la Piscine,
    38406 St-Martin d’H\`eres, France 
  \and
  Instituto Geogr\'afico Nacional, Centro de Desarrollos
  Tecnol\'ogicos, Observatorio de Yebes, Apartado 148,
  19080 Yebes, Spain    }
\date{Received June 29, 2021; accepted July 30, 2021}

\abstract{After the previous discovery of MgC$_3$N and MgC$_4$H in IRC\,+10216, a deeper
  Q-band (31.0-50.3 GHz) integration on this source
  had revealed two additional series of harmonically related doublets that we assigned 
  on the basis of quantum mechanical calculations 
  to the larger radicals MgC$_5$N and MgC$_6$H.
  The results presented here extend and confirm previous results on magnesium-bearing molecules in IRC\,+10216.
  We derived column densities of (4.7\,$\pm$\,1.3)\,$\times$\,10$^{12}$ for MgC$_5$N
    and (2.0\,$\pm$\,0.9)\,$\times$\,10$^{13}$ for MgC$_6$H, which imply that MgC$_5$N/MgC$_3$N\,=\,0.5
    and MgC$_6$H/MgC$_4$H\,=\,0.9. Therefore, MgC$_5$N and MgC$_6$H are present with column
      densities not so different from 
    those of the immediately shorter analogs. The synthesis
  of these large magnesium cyanides and acetylides in IRC\,+10216 can be explained for
  their shorter counterparts by a
  two-step process initiated by the radiative association of Mg$^+$ with large cyanopolyynes
  and polyynes, which are still quite abundant in this source, followed by the dissociative
  recombination of the ionic complexes.}

\keywords{molecular data ---  line: identification --- stars: carbon --- circumstellar matter ---
  stars: individual (IRC\,+10216)  --- astrochemistry}

\titlerunning{MgC$_5$N \& MgC$_6$H and IRC\,+10216}
\authorrunning{J. R. Pardo et al.}

\maketitle

\section{Introduction}

The vast majority of gas-phase metal-bearing molecules detected in space have been observed for the first time in
the carbon-rich circumstellar envelope (CSE) IRC\,+10216. It
is worth noting that atomic metals such as Na, K, Ca, Fe, Cr, and/or their cations are
found in the gas phase in IRC\,+10216 \citep{Mauron2010}, pointing toward a rich
metal chemistry in the outer CSE. The metal halides NaCl, AlCl, KCl, and AlF were the
first molecules to contain metals detected in this source and in space \citep{Cernicharo1987}.
Since these early detections, several new metal-bearing species
such as MgNC, MgCN, HMgNC, MgC$_3$N, MgCCH, MgC$_4$H, NaCN, SiCN, SiNC, AlNC, KCN, FeCN, and CaNC
have been detected \citep{Guelin1993,Guelin2000,Guelin2004,Kawaguchi1993,Turner1994,Ziurys1995,Ziurys2002,Pulliam2010,Zack2011,Cabezas2013,Agundez2014,Cernicharo2019a,Cernicharo2019b}. Among them, the family of metal
cyanide and isocyanide species is clearly the most numerous.

Most of the discoveries of these metal-bearing species in IRC\,+10216 have been done
through the interplay between laboratory rotational spectroscopy experiments and radioastronomical
observations. Nevertheless, the detection of further metal-containing
molecules is hampered by the lack of new spectroscopic data. These molecules are
transient species under terrestrial physical conditions and their generation
in laboratories is not trivial. In addition, their spectroscopic characterization
can be a 
difficult task because they usually have many electronic states
of different multiplicities with similar energies.

The improvements in sensitivity, spectral resolution, and bandwidth
achieved by telescopes and receivers in recent years allows one to perform very deep integrations
that are powerful
spectroscopic tools for analyzing the chemical composition of
astronomical objects. Once the lines coming from the isotopologues and vibrationally
excited states of already known molecules are identified, a forest of unidentified
lines emerges waiting to be assigned such as 
those registered in a laboratory spectrum. The
analysis of these spectral features, supported by high level ab initio calculations,
allows one to discover new molecules despite no laboratory data being available.
This procedure
has recently been employed in the astronomical discovery of MgC$_3$N and MgC$_4$H in
IRC\,+10216 \citep{Cernicharo2019b}, and of vibrationally excited HC$_7$N and
HC$_9$N \citep{Pardo2020} in this same source. 
The new sensitive broad band receivers used in this work have also allowed us to detect a
 large number of molecules in TMC-1, including pure hydrocarbon
  cycles such as indene and cyclopentadiene \citep{Cernicharo2021}.

In this letter we present the discovery in space of MgC$_5$N and MgC$_6$H, from
very long Q-band (31.0-50.3 GHz) integrations
toward IRC\,+10216 carried out with the Yebes 40m telescope, and precise quantum chemical calculations.
The observational setup is briefly summarized in section~\ref{sct:obs}. The bulk of this letter
is section~\ref{sct:hunt}, which is devoted to presenting what have been, up to now, two ladders of U-lines from our
observations, and their careful analysis aimed at identifying their molecular carriers and ruling
out other possible candidates. Although all of the doublets now assigned to MgC$_5$N ($N$\,=\,28-27 to $N$\,=\,41-40) are
well detected and the central frequencies
of their separate components can be well established, the same is not so easy for MgC$_6$H, even if it
could be more abundant, due to a much lower dipole moment.
Finally, in section~\ref{sec:chemistry} we discuss the
chemistry of magnesium in IRC\,+10216 after the discovery of these species, and we present our
conclusions in section~\ref{sct:concl}.

\section{Observations}
\label{sct:obs}

The observations presented in this letter are part of very deep integrations
from 31.0 to 50.3 GHz toward IRC\,+10216, which were carried out
from May 2019 to April 2021 with the 40 meter antenna
of the Centro Astron\'omico de Yebes (IGN, Spain) as part of the
Nanocosmos project\footnote{\texttt{https://nanocosmos.iff.csic.es/}}.
The experimental setup is described in
detail by \cite{Tercero2021} and further details can be found
in \cite{Pardo2020}. The most relevant information concerning the observations
leading to the results presented in this letter are described in
the reminder of this section.

The receiver consists of two HEMT cold amplifiers covering the
 31.0-50.3 GHz band with horizontal and vertical
polarizations. Receiver temperatures range from 22 K
at 31 GHz to 42 K at 50 GHz. The backends are 16$\times$2.5
GHz Fast Fourier Transform Spectrometers (FFTS), with a
spectral resolution of 38.1 kHz, although the data were smoothed
by a factor of six for their final presentation, resulting in a spectral resolution of 229 kHz.

The observations were achieved in position switching mode with
an off position at 300$"$ in azimuth. Pointing corrections were
obtained by observing the strong SiO masers of R\,Leo and
were always within 2-3$"$. The intensity scale provided in the
figures of this letter was antenna temperature (T$_A^*$) corrected
for atmospheric absorption using the ATM package
\citep{Cernicharo1985,Pardo2001}. Calibration
uncertainties are estimated to be within 10~\%.
The final noise level obtained, as a function of frequency, is plotted in Figure \ref{fg:mgcxn}.

\begin{table*}
  \caption{Calculated and observational parameters of MgC$_5$N and MgC$_6$H lines in our Q-band observations
    of IRC\,+10216. Quantum numbers are  N(J) for both species. Frequencies
    are given in MHz, energies in K, W=$\int$T$_{MB}$dv in mK$\cdot$kms$^{-1}$.
    We note that
    R is the rotational diagram = $ln(3\;k_B\;W / 8\;\pi^3\;\nu_{rest}\;S_{ul}\;\mu^2)$. The symbol = in the observed
    frequencies means that the fitting was forced to use the calculated frequency resulting from all the other observed
    frequencies that could be freely fitted from the data. Blanks are left for observed parameters that could not be obtained
    due to blending with much stronger lines or frequencies outside the survey's range. The boldface indicates lines shown
    in Figure \ref{fg:mgc5n}.}
\label{tb:mgc5n}
\centering
\begin{tabular}{ccccccccccccc}

& \multicolumn{6}{c}{MgC$_5$N} & \multicolumn{6}{c}{MgC$_6$H} \\
\cmidrule(lr){2-7} \cmidrule(lr){8-13}
{\small N(J)   }&{\small  $\nu_{calc}$ }&{\small   $\nu_{obs}$ }&{\small  E$_{up}$ }&{\small  S$_{ul}$ }&{\small  W    }&{\small  R   }&{\small  $\nu_{calc}$ }&{\small   $\nu_{obs}$ }&{\small  E$_{up}$ }&{\small  S$_{ul}$ }&{\small  W    }&{\small  R }\\
\hline
{\small  27(53/2)$\rightarrow$26(51/2)  }&{\small             }&{\small            }&{\small          }&{\small          }&{\small         }&{\small           }&{\small    31305.40 }&{\small    31305.41 }&{\small     20.9 }&{\small     26.5 }&{\small     332 }&{\small      23.094     }\\
{\small  27(55/2)$\rightarrow$26(53/2)  }&{\small             }&{\small            }&{\small          }&{\small          }&{\small         }&{\small           }&{\small    31307.36 }&{\small    31307.36 }&{\small     20.9 }&{\small     27.5 }&{\small     332 }&{\small      23.056     }\\
{\small  28(55/2)$\rightarrow$27(53/2)  }&{\small  {\bf  32278.59} }&{\small   {\bf 32278.52} }&{\small     {\bf 22.5} }&{\small   {\bf   27.5} }&{\small   {\bf   152} }&{\small  {\bf  20.098}  }&{\small    {\bf 32464.88} }&{\small    {\bf 32464.93} }&{\small    {\bf   22.5} }&{\small    {\bf   27.5} }&{\small     {\bf  223} }&{\small     {\bf   22.623}     }\\
{\small  28(57/2)$\rightarrow$27(55/2)  }&{\small  {\bf  32280.40} }&{\small   {\bf 32280.53} }&{\small     {\bf 22.5} }&{\small   {\bf   28.5} }&{\small    {\bf  379} }&{\small   {\bf 20.978}  }&{\small    {\bf 32466.84} }&{\small    {\bf   =     } }&{\small     {\bf  22.5} }&{\small     {\bf  28.5} }&{\small     {\bf   74} }&{\small      {\bf  21.489}     }\\
{\small  29(57/2)$\rightarrow$28(55/2)  }&{\small  {\bf   33431.38} }&{\small   {\bf 33431.36} }&{\small     {\bf 24.1} }&{\small  {\bf    28.5} }&{\small   {\bf   273} }&{\small  {\bf  20.616}  }&{\small    33624.36 }&{\small       =     }&{\small     24.1 }&{\small     28.5 }&{\small      67 }&{\small      21.351     }\\
{\small  29(59/2)$\rightarrow$28(57/2)  }&{\small  {\bf   33433.19} }&{\small   {\bf 33433.20} }&{\small     {\bf 24.1} }&{\small   {\bf   29.5} }&{\small   {\bf   273} }&{\small   {\bf 20.582}  }&{\small    33626.32 }&{\small    33626.31 }&{\small     24.1 }&{\small     29.5 }&{\small      67 }&{\small      21.316     }\\
{\small  30(59/2)$\rightarrow$29(57/2)  }&{\small    34584.17 }&{\small   34584.20 }&{\small     25.7 }&{\small     29.5 }&{\small     310 }&{\small   20.673  }&{\small    {\bf 34783.84} }&{\small   {\bf 34784.02} }&{\small    {\bf 25.7} }&{\small   {\bf  29.5} }&{\small    {\bf 244} }&{\small    {\bf  22.571}    }\\
{\small  30(61/2)$\rightarrow$29(59/2)  }&{\small    34585.98 }&{\small   34586.10 }&{\small     25.7 }&{\small     30.5 }&{\small     433 }&{\small   20.976  }&{\small    {\bf 34785.79} }&{\small   {\bf 34785.76} }&{\small    {\bf 25.7} }&{\small   {\bf  30.5} }&{\small    {\bf 122} }&{\small    {\bf  21.844}     }\\
{\small  31(61/2)$\rightarrow$30(59/2)  }&{\small  {\bf   35736.95} }&{\small   {\bf 35736.93} }&{\small    {\bf  27.4} }&{\small   {\bf   30.5} }&{\small    {\bf  282} }&{\small  {\bf  20.514}  }&{\small    35943.31 }&{\small    35943.32 }&{\small     27.4 }&{\small     30.5 }&{\small     111 }&{\small      21.719     }\\
{\small  31(63/2)$\rightarrow$30(61/2)  }&{\small  {\bf   35738.76} }&{\small   {\bf 35738.74} }&{\small    {\bf  27.4} }&{\small   {\bf   31.5} }&{\small    {\bf  169} }&{\small  {\bf  19.971}  }&{\small    35945.27 }&{\small    35945.15 }&{\small     27.4 }&{\small     31.5 }&{\small     167 }&{\small      22.092     }\\
{\small  32(63/2)$\rightarrow$31(61/2)  }&{\small    36889.72 }&{\small   36889.68 }&{\small     29.2 }&{\small     31.5 }&{\small     155 }&{\small   19.851  }&{\small    {\bf 37102.79} }&{\small    {\bf   =     } }&{\small   {\bf  29.2} }&{\small   {\bf  31.5} }&{\small  {\bf   102} }&{\small   {\bf   21.567}     }\\
{\small  32(65/2)$\rightarrow$31(63/2)  }&{\small    36891.53 }&{\small   36891.22 }&{\small     29.2 }&{\small     32.5 }&{\small     155 }&{\small   19.820  }&{\small    {\bf 37104.74} }&{\small    {\bf 37104.61} }&{\small   {\bf  29.2} }&{\small   {\bf  32.5} }&{\small  {\bf   305} }&{\small   {\bf   22.634}     }\\
{\small  33(65/2)$\rightarrow$32(63/2)  }&{\small  {\bf   38042.49} }&{\small   {\bf 38042.49} }&{\small    {\bf  31.0} }&{\small   {\bf   32.5} }&{\small   {\bf   143} }&{\small   {\bf 19.705}  }&{\small    {\bf 38262.26} }&{\small    {\bf 38262.21} }&{\small   {\bf  31.0} }&{\small   {\bf  32.5} }&{\small  {\bf    94} }&{\small   {\bf   21.422}     }\\
{\small  33(67/2)$\rightarrow$32(65/2)  }&{\small  {\bf   38044.30} }&{\small   {\bf 38044.36} }&{\small    {\bf  31.0} }&{\small   {\bf   33.5} }&{\small   {\bf   285} }&{\small   {\bf 20.368}  }&{\small    {\bf 38264.21} }&{\small    {\bf   =     } }&{\small   {\bf  31.0} }&{\small   {\bf  33.5} }&{\small  {\bf   140} }&{\small   {\bf   21.797}     }\\
{\small  34(67/2)$\rightarrow$33(65/2)  }&{\small    39195.26 }&{\small   39195.18 }&{\small     32.9 }&{\small     33.5 }&{\small     220 }&{\small   20.077  }&{\small    39421.72 }&{\small       =     }&{\small     32.9 }&{\small     33.5 }&{\small     173 }&{\small      21.976     }\\
{\small  34(69/2)$\rightarrow$33(67/2)  }&{\small    39197.06 }&{\small   39197.00 }&{\small     32.9 }&{\small     34.5 }&{\small     307 }&{\small   20.384  }&{\small    39423.68 }&{\small    39423.39 }&{\small     32.9 }&{\small     34.5 }&{\small     346 }&{\small      22.640     }\\
{\small  35(69/2)$\rightarrow$34(67/2)  }&{\small    40348.01 }&{\small   40347.94 }&{\small     34.9 }&{\small     34.5 }&{\small     285 }&{\small   20.280  }&{\small    40581.19 }&{\small             }&{\small     34.9 }&{\small     34.5 }&{\small         }&{\small                 }\\
{\small  35(71/2)$\rightarrow$34(69/2)  }&{\small    40349.82 }&{\small   40349.69 }&{\small     34.9 }&{\small     35.5 }&{\small     285 }&{\small   20.251  }&{\small    40583.15 }&{\small       =     }&{\small     34.9 }&{\small     35.5 }&{\small     241 }&{\small      22.220     }\\
{\small  36(71/2)$\rightarrow$35(69/2)  }&{\small  {\bf   41500.77} }&{\small   {\bf 41500.78} }&{\small    {\bf  36.8} }&{\small   {\bf   35.5} }&{\small  {\bf    266} }&{\small  {\bf  20.152}  }&{\small    {\bf 41740.66} }&{\small    {\bf 41740.10} }&{\small  {\bf   36.8} }&{\small   {\bf  35.5} }&{\small  {\bf   150} }&{\small  {\bf    21.715}     }\\
{\small  36(73/2)$\rightarrow$35(71/2)  }&{\small  {\bf   41502.57} }&{\small   {\bf 41502.43} }&{\small    {\bf  36.9} }&{\small   {\bf   36.5} }&{\small  {\bf    152} }&{\small  {\bf  19.564}  }&{\small    {\bf 41742.61} }&{\small    {\bf    =    } }&{\small  {\bf   36.9} }&{\small   {\bf  36.5} }&{\small  {\bf   112} }&{\small   {\bf   21.400}     }\\
{\small  37(73/2)$\rightarrow$36(71/2)  }&{\small  {\bf   42653.51} }&{\small   {\bf 42653.55} }&{\small    {\bf  38.9} }&{\small   {\bf   36.5} }&{\small  {\bf    177} }&{\small  {\bf  19.693}  }&{\small    {\bf 42900.12} }&{\small    {\bf 42900.43} }&{\small  {\bf   38.9} }&{\small   {\bf  36.5} }&{\small  {\bf   210} }&{\small  {\bf    21.999}     }\\
{\small  37(75/2)$\rightarrow$36(73/2)  }&{\small  {\bf   42655.32} }&{\small   {\bf 42655.25} }&{\small    {\bf  38.9} }&{\small   {\bf   37.5} }&{\small  {\bf    177} }&{\small  {\bf  19.666}  }&{\small    {\bf 42902.07} }&{\small    {\bf 42901.47} }&{\small  {\bf   38.9} }&{\small   {\bf  37.5} }&{\small  {\bf    70} }&{\small  {\bf    20.873}     }\\
{\small  38(75/2)$\rightarrow$37(73/2)  }&{\small    43806.25 }&{\small   43806.45 }&{\small     41.0 }&{\small     37.5 }&{\small     100 }&{\small   19.065  }&{\small    44059.20 }&{\small             }&{\small     41.0 }&{\small     37.5 }&{\small         }&{\small                 }\\
{\small  38(77/2)$\rightarrow$37(75/2)  }&{\small    43808.06 }&{\small   43807.98 }&{\small     41.0 }&{\small     38.5 }&{\small     200 }&{\small   19.731  }&{\small    44061.52 }&{\small    44061.53 }&{\small     41.0 }&{\small     38.5 }&{\small      66 }&{\small      20.757     }\\
{\small  39(77/2)$\rightarrow$38(75/2)  }&{\small    44958.98 }&{\small   44958.88 }&{\small     43.2 }&{\small     38.5 }&{\small      94 }&{\small   18.952  }&{\small    45219.04 }&{\small             }&{\small     43.2 }&{\small     38.5 }&{\small         }&{\small                 }\\
{\small  39(79/2)$\rightarrow$38(77/2)  }&{\small    44960.79 }&{\small   44960.46 }&{\small     43.2 }&{\small     39.5 }&{\small     157 }&{\small   19.437  }&{\small    45220.99 }&{\small             }&{\small     43.2 }&{\small     39.5 }&{\small         }&{\small                 }\\
{\small  40(79/2)$\rightarrow$39(77/2)  }&{\small    46111.71 }&{\small   46111.60 }&{\small     45.4 }&{\small     39.5 }&{\small     118 }&{\small   19.132  }&{\small    46378.49 }&{\small       =     }&{\small     45.4 }&{\small     39.5 }&{\small     146 }&{\small      21.480     }\\
{\small  40(81/2)$\rightarrow$39(79/2)  }&{\small    46113.51 }&{\small   46113.74 }&{\small     45.4 }&{\small     40.5 }&{\small     178 }&{\small   19.512  }&{\small    46380.45 }&{\small             }&{\small     45.4 }&{\small     40.5 }&{\small         }&{\small                 }\\
{\small  41(81/2)$\rightarrow$40(79/2)  }&{\small    47264.42 }&{\small   47264.57 }&{\small     47.6 }&{\small     40.5 }&{\small     168 }&{\small   19.434  }&{\small    47537.94 }&{\small             }&{\small     47.6 }&{\small     40.5 }&{\small         }&{\small                 }\\
{\small  41(83/2)$\rightarrow$40(81/2)  }&{\small    47266.23 }&{\small   47266.21 }&{\small     47.6 }&{\small     41.5 }&{\small      84 }&{\small   18.716  }&{\small    47539.90 }&{\small             }&{\small     47.6 }&{\small     41.5 }&{\small         }&{\small                 }\\
{\small  42(83/2)$\rightarrow$41(81/2)  }&{\small    48417.14 }&{\small            }&{\small     50.0 }&{\small     41.5 }&{\small         }&{\small           }&{\small    48697.40 }&{\small    48697.07 }&{\small     50.0 }&{\small     41.5 }&{\small     158 }&{\small      21.460     }\\
{\small  42(85/2)$\rightarrow$41(83/2)  }&{\small    48418.94 }&{\small            }&{\small     50.0 }&{\small     42.5 }&{\small         }&{\small           }&{\small    48699.35 }&{\small             }&{\small     50.0 }&{\small     42.5 }&{\small         }&{\small                 }\\
\hline
\end{tabular}
\end{table*}

\section{Hunting new Magnesium radicals in IRC\,+10216}
\label{sct:hunt}

The final data product of our IRC\,+10216 Q-band survey consists of a 229 kHz spectrum
with line fit curves and tables with the parameters of those fits and the list of
molecular transitions to which they belong. The complete survey will be published shortly
in a separate paper \citep{Pardo2021}. The line identification process has been
done in several steps with the weakest features sometimes being confused with baseline
uncertainties and needing a careful revisit of the raw data.
We have used the MADEX catalog \citep{Cernicharo2012}, 
the Cologne Database of Molecular Spectroscopy catalog (CDMS; \citealt{Muller2005}), and
the JPL catalog \citep{Pickett1998}. The list of frequencies
for the remaining U-lines at a given stage of the analysis was carefully checked to
search for harmonic relations that could lead to finding new species.

\begin{figure}[t]
\includegraphics[width=\columnwidth]{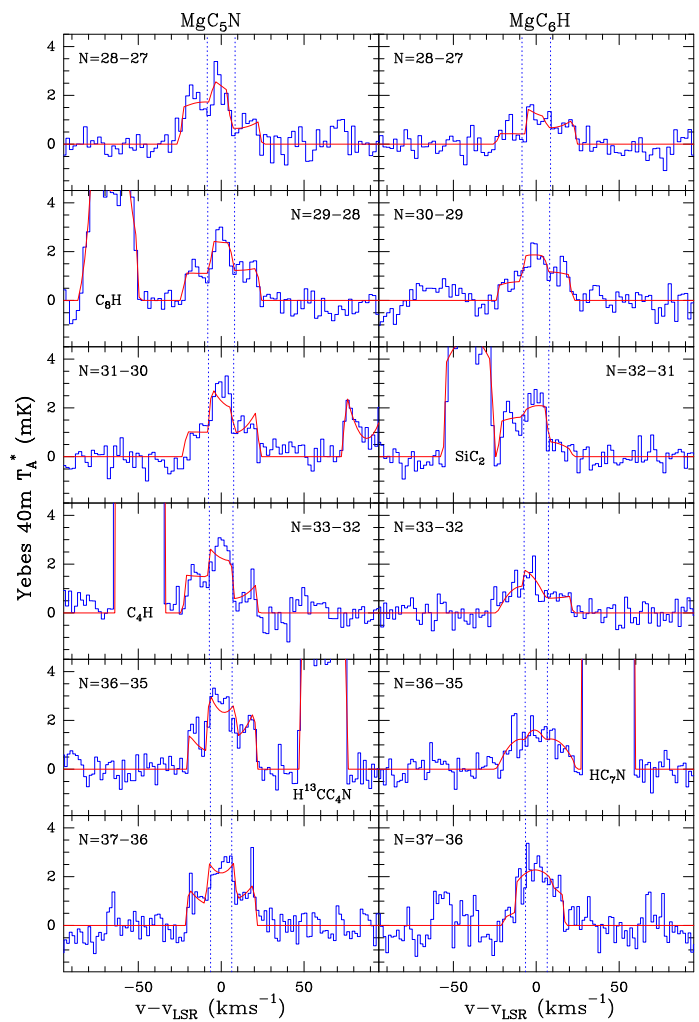}
\caption{
  Selection of the doublets corresponding to MgC$_5$N
  and MgC$_6$H as they appear in the IRC\,+10216
  ultra-deep Q-band survey carried out with
  the Yebes 40m telescope. Data: Blue histograms. Model fit: Red curves.
  Calculated central frequencies of the doublets: Blue dotted lines.}
\label{fg:mgc5n} \end{figure}

\begin{figure}[t]
\includegraphics[width=\columnwidth]{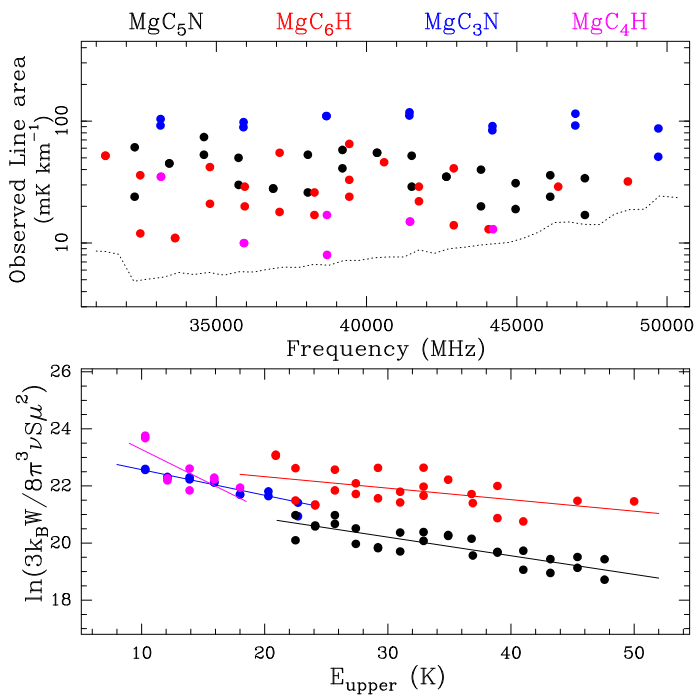}
\caption{Observed integrated line areas of MgC$_3$N, MgC$_4$H, MgC$_5$N, and MgC$_6$H toward IRC+10216 in the Q-band,
  and the rotational diagram for each
  molecule built from those integrated areas, together with the best fits that provide an estimation
  of their column densities and rotational temperatures. The dotted curve in the top panel is the noise
  level achieved in our survey across the whole frequency range. A few lines beyond 44 GHz are below the detectability
  limit and for this reason they do not appear in Table \ref{tb:mgc5n}.}
\label{fg:mgcxn} \end{figure}

After our publications on the identification of a series of lines belonging to
MgC$_3$N and MgC$_4$H \citep{Cernicharo2019b} and, later, on
the $\nu_{19}$ mode of HC$_9$N \citep{Pardo2020},  
 other sets of
spectral features, with the appearance of doublets, got our attention. 
All of those doublets have intensities below $\sim$3 mK at the final resolution of the data set.
Although they are among the weakest features arising above the noise level achieved in our survey, they can
be fitted using the GILDAS Shell method to two separate components having the 14.5 km s$^{-1}$ expansion
velocity characteristic of the source.
Table \ref{tb:mgc5n}
summarizes the line parameters, whereas a selection of observed lines (and their fits) is shown in Fig.~\ref{fg:mgc5n}. Column
densities and rotational temperatures were derived from rotation diagrams (see Fig.~\ref{fg:mgcxn})
using the data given in Table~\ref{tb:mgc5n}. The results are the following:
$T_{rot}$\,=\,15.4\,$\pm$\,1.8 K and $N_{col}$\,=\,(4.7\,$\pm$\,1.3)\,$\times$\,10$^{12}$ for MgC$_5$N
and $T_{rot}$\,=\,24.8\,$\pm$\,8.9 K and $N_{col}$\,=\,(2.0\,$\pm$\,0.9)\,$\times$\,10$^{13}$ for MgC$_6$H.

Even if the column density is larger for MgC$_6$H than for MgC$_5$N, as it was found for MgC$_4$H versus
MgC$_3$N \citep{Cernicharo2019b}, the detection of MgC$_5$N is clearer
 because its lines are strong enough and well detected in a reliable way throughout almost the whole rotational
 ladder covered by our data ($N$\,=\,28-27 to $N$\,=\,41-40, with only $N$\,=\,42-41 being below the noise level). The
much larger dipole moment of
MgC$_5$N (7.3 Debye) with respect to MgC$_6$H (2.5 Debye), as discussed
later in this section, is responsible for that. In
fact, the assignment to MgC$_6$H of the weak features shown in the right column of Fig.~\ref{fg:mgc5n} (the most
prominent in our data through its covered rotational ladder, $N$\,=\,27-26 to $N$\,=\,42-41) has been possible only
after extra observations were scheduled and achieved in 2021. In any case, it is worth noting that for MgC$_5$N, we derived
a lower rotational temperature than for MgC$_6$H, which is consistent with the higher dipole moment of the cyanide compared
to the acetylide. The reminder of this section is devoted to the detailed quantum mechanical calculations
carried out to do the assignments and to explore other alternatives for the carriers.

In order to obtain precise geometries and spectroscopic molecular parameters that help in the assignment
of the observed lines, we carried out high-level ab initio calculations for the following two possible
candidates: MgC$_5$N and MgC$_6$H. All the calculations were performed using the
Molpro 2020.2 \citep{Werner2020} and Gaussian09 \citep{Frisch2013} program packages. The geometry
optimization calculations were carried out at the CCSD(T)-F12/cc-pCVTZ-F12 level of
theory \citep{Raghavachari1989,Adler2007,Knizia2009,Hill2010a,Hill2010b}, which has been proved as
a suitable method to accurately reproduce the molecular geometry of analog molecules \citep{Cernicharo2019b}.
We followed the same strategy previously used by us for the related molecular systems MgC$_3$N and
MgC$_4$H \citep{Cernicharo2019b}. In this manner, we scaled our calculations for MgC$_5$N and MgC$_6$H
using the experimental and theoretical results of MgC$_3$N and MgC$_4$H, respectively.

According to our calculations, the electronic ground state for MgC$_5$N and MgC$_6$H is $^2\Sigma$. The
calculated spectroscopic parameters for both species are shown in Table \ref{abini}, together with those
for MgC$_3$N and MgC$_4$H. As expected, the $B$, $D_N$, and $\gamma$ values for MgC$_5$N and MgC$_6$H
are quite similar. However, large differences were found between the predicted dipole moments, being
7.3 D and 2.5 D for MgC$_5$N and MgC$_6$H, respectively. A comparison of the scaled values with those
obtained from fitting the IRC\,+10216 lines allowed us to assign the two series of lines to MgC$_5$N and
MgC$_6$H. The relative errors for the $B$ constant are 0.024\% and 0.017\% for MgC$_5$N and MgC$_6$H,
respectively. For $\gamma$ values, the errors are larger, 2.8\% and 4.1\% for MgC$_5$N and MgC$_6$H,
respectively. Similar discrepancies were found for MgC$_3$N and MgC$_4$H (see Table \ref{abini}). For
$D_N$, the agreement between calculated and determined values for MgC$_5$N is very good, within around 4\%.
However, for MgC$_6$H, the $D_N$ value was not determined from our fit and, thus, a
comparison with the ab initio values makes no sense.

  As the doublets of MgC$_5$N are directly unquestionable from the data, those of MgC$_6$H, due to
  a lower signal-to-noise ratio, require further support in particular to rule out $^1\Sigma$ species.
  In fact, there are not many alternatives to MgC$_6$H, basically the C$_7$N radical and the
  C$_7$N$^-$ anion as their theoretical rotational constants $B$ are 583.1 and 582.0 MHz,
  respectively \citep{Botschwina1997,Botschwina2008}.

  The anion C$_7$N$^-$ is predicted to have a $^1\Sigma$ electronic ground state \citep{Botschwina2008}
  and thus it could be an alternative to MgC$_6$H. However, even if a lower noise level should be
  desirable, the observed line widths are too large for a $^1\Sigma$ species. Moreover, the
  C$_7$N radical itself should be present in our data, but we have no evidence of lines related by half integer
  quantum numbers that could be assigned to it as its electronic ground state  was calculated to
  be $^2\Pi$ \citep{Botschwina1997}.

\begin{table*}
  \caption{Spectroscopic parameters $B$, $D_N$, and $\gamma$ for MgC$_3$N, MgC$_5$N, MgC$_4$H, and MgC$_6$H (all in MHz).}
\label{abini}
\centering
\begin{tabular}{{lcccccc}}
\hline
\hline
 & & & & & & \\
&\multicolumn{2}{c}{MgC$_3$N}&\multicolumn{3}{c}{MgC$_5$N} \\
\cmidrule(lr){2-3} \cmidrule(lr){4-6}
Parameter & Calc.\tablefootmark{a} & Exp.\tablefootmark{b} & Calc.\tablefootmark{a} & Scaled\tablefootmark{c} & Exp.\tablefootmark{d}\\
\hline
$B$            &    1376.50                &     1380.888               &  574.74                 &  576.57                &  576.43106(125)     \\
$D_N$          &  0.0640$\times$$10^{-3}$  &   0.0760$\times$$10^{-3}$  &  5.87$\times$$10^{-6}$  &  6.98$\times$$10^{-6}$ &    7.325(483) $\times$$10^{-6}$  \\
$\gamma$       &    4.09                   &     4.35                   &  1.64                   &  1.75                                                            &    1.8071(386)      \\                              
 & & & & & & \\
\hline
\hline
 & & & & & & \\
&\multicolumn{2}{c}{MgC$_4$H}&\multicolumn{3}{c}{MgC$_6$H} \\
\cmidrule(lr){2-3} \cmidrule(lr){4-6}
Parameter & Calc.\tablefootmark{a} & Exp.\tablefootmark{b} & Calc.\tablefootmark{a} & Scaled\tablefootmark{c} & Exp.\tablefootmark{d}\\
\hline
$B$            &   1377.38                 &     1381.512                 &  578.11                     &  579.85                  &   579.75150(459)     \\
$D_N$          &  0.0601$\times$$10^{-3}$  &   0.074$\times$$10^{-3}$   &  5.83$\times$$10^{-6}$        &  7.02$\times$$10^{-6}$   &     2.53(214)    $\times$$10^{-6}$  \\
$\gamma$       &    4.43                   &     4.7                      &  1.76                       &  1.87                    &     1.9551(141)      \\                                                                                                                                  
 & & & & & & \\
\hline
\hline
\end{tabular}
\tablefoot{\\
  \tablefoottext{a}{This work. The $B$ rotational constant was calculated at the CCSD(T)-F12/cc-pCVTZ-F12 level of
    theory, while $D_N$ and $\gamma$ constants were calculated at the MP2/cc-pVTZ level of theory.} \\
\tablefoottext{b}{\citet{Cernicharo2019b}.} \\
\tablefoottext{c}{This work; scaled by the ratio Exp/Calc. of the corresponding parameter for MgC$_3$N and MgC$_4$H species, see text}.\\
\tablefoottext{d}{This work.}
\\
}
\end{table*}

\section{Magnesium chemistry in IRC\,+10216}
\label{sec:chemistry}

The formation of MgC$_5$N and MgC$_6$H in IRC\,+10216 occurs very likely through the same mechanism that forms other smaller magnesium cyanides and acetylides previously detected in this same source, that is, MgNC, MgCN, HMgNC, MgC$_3$N, MgCCH, and MgC$_4$H. Based on the work of \cite{Petrie1996}, the synthesis of these types of molecules is thought to proceed through a two-step gas-phase mechanism in which large cyanopolyynes and polyynes associate radiatively with Mg$^+$ and then the Mg$^+$/NC$_{2n+1}$H and Mg$^+$/C$_{2n}$H$_2$ complexes fragment upon dissociative recombination with electrons \citep{Millar2008,Cabezas2013,Cernicharo2019b}.

Here we have expanded the chemical network used in \cite{Cernicharo2019b} to include the larger species MgC$_5$N and MgC$_6$H. The chemical model of the outer envelope is based on \cite{Agundez2017}. As in \cite{Cernicharo2019b}, the rate coefficients for the radiative associations of Mg$^+$ with polyynes and cyanopolyynes are taken from \cite{Dunbar2002} and we assume that open shell metal-containing species (MgNC, MgCN, MgC$_3$N, MgC$_5$N, MgCCH, MgC$_4$H, and MgC$_6$H) react with electrons and H atoms, while closed shell molecules (HMgNC being the only one considered) do not react. The ion complexes Mg$^+$/NC$_{2n+1}$H and Mg$^+$/C$_{2n}$H$_2$ may have multiple channels of fragmentation when reacting with electrons. The branching ratios of each channel are unknown, although they are key since they are the main parameters that set the relative abundances between the different magnesium cyanide derivatives and between the various magnesium acetylides. As in \cite{Cernicharo2019b}, we adjusted these branching ratios to reproduce the relative abundances observed.

  First of all, if we look at
the magnesium cyanides,
  we find that their
  observed relative abundances
  follow this sequence: 
MgNC/HMgNC/MgCN/MgC$_3$N/MgC$_5$N\,$\sim$\,1/0.05/0.06/0.7/0.35 (\citealt{Cabezas2013,Cernicharo2019b}; this work). We find that to reproduce these ratios, the branching ratios of the channels yielding MgCN, HMgNC, MgC$_3$N, and MgC$_5$N must be 0.05, 0.01, 0.7, and 0.35 of that yielding MgNC. These values, which are similar to those given in \cite{Cernicharo2019b}, roughly correspond to the column density ratios observed, except for HMgNC, which in the chemical model is assumed to have a slower destruction rate due to its closed shell nature.

\begin{figure}
\includegraphics[width=\columnwidth]{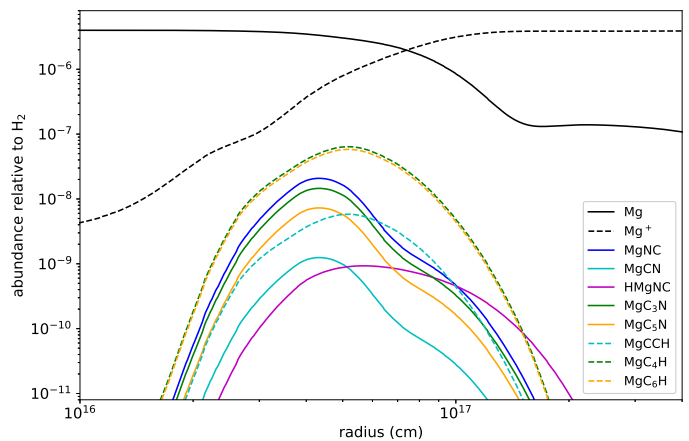}
\caption{Abundances calculated with the chemical model for Mg, Mg$^+$, and all Mg-bearing molecules detected in IRC\,+10216.}
\label{fig:model}
\end{figure}

For the magnesium acetylides, the observed column density ratios are MgCCH/MgC$_4$H/MgC$_6$H\,$\sim$\,1/11/10 and we find that the branching ratios of the channels yielding MgC$_4$H and MgC$_6$H must be 11 and 10 times of that yielding MgCCH, which precisely follow the observed relative abundances since the three acetylides are open shell species and thus are assumed to have similar destruction rates. \cite{Mauron2010} did not observe atomic Mg in the outer layers of IRC\,+10216 and thus the abundance of atomic Mg is not well constrained. Here we adopted an initial abundance of Mg relative to H of 2\,$\times$10$^{-6}$, which accounts for about 5\,\% of the cosmic abundance of Mg and allowed us to reproduce the absolute column densities of the magnesium cyanides and acetylides observed. For example, the calculated column density of MgC$_5$N is 2.8\,$\times$\,10$^{12}$ cm$^{-2}$, while for MgC$_6$H we calculated 2.6\,$\times$\,10$^{13}$ cm$^{-2}$, which are only slightly below and above the values derived from the observations, respectively. The results from the chemical model are shown in Fig.~\ref{fig:model}.

It is interesting to note that MgNC should be significantly favored in the dissociative recombination of Mg$^+$/NC$_{2n+1}$H complexes, compared to the larger magnesium cyanides MgC$_3$N and MgC$_5$N. However, in the case of Mg$^+$/C$_{2n}$H$_2$ complexes, the larger acetylides MgC$_4$H and MgC$_6$H should be strongly favored compared to MgCCH. It would be interesting to verify this point, which directly arises from the observed relative abundances in IRC\,+10216, by studying the dissociative recombination of these complexes. If large fragments tend to be favored, it would not be surprising that even larger members of the series of magnesium cyanides and acetylides such as MgC$_7$N and MgC$_8$H could be present in IRC\,+10216 with abundances comparable to those of MgC$_5$N and MgC$_6$H, respectively.

\section{Conclusions} \label{sct:concl}
  This work has extended the catalog of molecular species found for the first time in IRC+10216, and in space,
  with the following two new members: MgC$_5$N and MgC$_6$H. But, more interestingly, their column densities have been found
  to be not so different from their shorter counterparts so that it could be possible to find longer
  magnesium cyanides and acetylides in IRC\,+10216. This and other recent results from our group suggest
  that new receivers covering the $\sim$23-31 GHz range
  should be a major priority for the Yebes 40m radiotelescope, as they could lead to new and
  interesting findings of heavy molecules in objects such as IRC+10216 or TMC-1 that would have
  an important impact on our understanding of astrochemistry.

  A forthcoming paper will present the results of the whole Yebes IRC+10216 Q-band survey. In it, the
  number of U-lines has been drastically reduced due to the progress made in line catalogs thanks to
  works as this one, with two molecular species being spectroscopically characterized from astronomical
  data without any previous laboratory data.

\begin{acknowledgements}
  We thank Ministerio de Ciencia e Innovación of Spain (MI-CIU) for
  funding support through projects AYA2016-75066-C2-1-P, PID2019-
106110GB-I00, PID2019-107115GB-C21 / AEI / 10.13039/501100011033, and
PID2019-106235GB-I00. We also thank ERC for funding through grant ERC-
2013-Syg-610256-NANOCOSMOS. M.A. thanks MICIU for grant RyC-2014-
16277.
\end{acknowledgements}

\end{document}